\documentclass[journal]{IEEEtran}
\usepackage{graphicx}
\usepackage{amssymb,amsmath}
\usepackage{epsf}
\usepackage{mathrsfs}

\newtheorem{theorem}{Theorem}[section]
\newtheorem{proposition}[theorem]{Proposition}

\newtheorem{lemma}[theorem]{Lemma}

\newtheorem{corollary}[theorem]{Corollary}

\def\R{\mathbb{R}}

\newcommand{\be}{\begin{eqnarray}}
\newcommand{\ee}{\end{eqnarray}}

\newcommand\abs[1]{{\left| {#1} \right|}}

\newcommand\norm[1]{{\| #1 \|}}

\newcommand\ket[1]{{ |{#1} \rangle }}
\newcommand\bra[1]{{ \langle {#1} | }}
\newcommand\braket[2]{{ \langle {#1} | {#2} \rangle}}

\newcommand{\snote}[1]{}
\newcommand{\jnote}[1]{}
\newcommand{\enote}[1]{}
\newcommand{\ignore}[1]{}

\DeclareMathOperator{\Coker}{Coker}
\DeclareMathOperator{\Ker}{Ker}
\DeclareMathOperator{\rank}{rank}
\newcommand{\sgn}{\mathrm{sgn}}

\newcommand{\eps}{\varepsilon}
\renewcommand{\epsilon}{\varepsilon}

\begin{document}
%
\title{\bf Generalized Performance of Concatenated Quantum Codes\\ -- a dynamical systems approach}
%
%
\author{Jesse~Fern,
        Julia~Kempe,
        Slobodan~N.~Simi\'c,
        Shankar~Sastry
\thanks{J. Fern is with the Department of Mathematics, UC Berkeley, CA, 94720-3840, E-mail: \texttt{jesse@math.berkeley.edu}}
\thanks{J. Kempe is with the Computer Science Division and Department of Chemistry UC Berkeley, CA, 94720, and also with CNRS-LRI UMR 8623, Universit\'e de Paris-Sud, 91405 Orsay, France}
\thanks{S. N. Simi\'c was with the Department of Electrical Engineering and Computer Science UC Berkeley, CA, 94720-1770. He is now with the Department of Mathematics, San Jos\'e State University, San Jos\'e, CA, 95192-0103, E-mail: \texttt{simic@math.sjsu.edu}}
\thanks {S. Sastry is with the Department of Electrical Engineering and Computer Science, UC Berkeley, CA, 94720-1770}}
\maketitle

\begin{abstract}
    We apply a dynamical systems approach to concatenation of quantum
  error correcting codes, extending and generalizing the results of
  Rahn et al. \cite{Rahn:02a} to both diagonal and non-diagonal
  channels. Our point of view is global: instead of focusing on
  particular types of noise channels, we study the geometry of the
  coding map as a discrete-time dynamical system on the entire space
  of noise channels.

  In the case of diagonal channels, we show that any code with
  distance at least three corrects (in the infinite concatenation
  limit) an open set of errors. For Calderbank-Shor-Steane (CSS) codes, we give a more precise
  characterization of that set. We show how to incorporate noise in the gates, thus completing the framework. We derive some general bounds for noise channels, which allows us to analyze several codes in detail.
\end{abstract}

\begin{keywords}
Quantum error correction, quantum channels,  quantum fault tolerance
\end{keywords}

%
\IEEEpeerreviewmaketitle

\section{Introduction}


In this paper we analyze quantum codes in essence, abstracting their
details as codes and extracting their fault tolerance properties using
a dynamical systems approach. This framework has been initiated by
Rahn et al. \cite{Rahn:02a}. They show how to incorporate diagonal
noise on the qubit into an {\em effective channel} on the {\em
  logical} qubits.

We broaden this viewpoint and extend their approach in several ways.
We look at the effective channel from a dynamical systems point of
view, using tools and methods from this field. In particular we
characterize the region of correctable errors using tools from the
analysis of fixed points and show how to incorporate perturbations of
the coding map.

Our second chain of results extends the results of \cite{Rahn:02a} to
the realistic model of faulty gates and general channels. Rahn et al.
only analyzed the depolarizing channel on the physical qubits as the
single source of noise. We show that incorporating noisy gates gives
rise to a {\em perturbed} effective channel. We also analyze general
noise on the qubits and give several bounds for the convergence of
non-diagonal channels to diagonal channels. Our results are supported
by several examples for the family of CSS-codes, which is the encoding
predominantly proposed for fault-tolerant quantum computing. We
simplify our bounds in the case of CSS codes and analyze the
$[[7,1,3]]$ code, the smallest member of the CSS family, in great
detail.

\paragraph{Structure of the paper}

We first introduce the dynamical systems approach in Section
\ref{sec:frame} and establish the notation and some basics. In Section
\ref{sec:diagonal} we extend this approach to diagonal channels,
including an analysis of regions of convergence.
Section~\ref{sec:faulty} deals with faulty gates.  In Section
\ref{sec:channels} we establish several results and examples for
non-diagonal (i.e. general) noise channels and in Section \ref{sec:canon}, we discuss a way to improve channels. Our approach allows to
drastically reduce the number of parameters, lending Quantum error correcting codes to an
elegant analysis. This however comes at some price, and in Section
\ref{sec:conclude} we outline some of the shortcomings of this
approach, before concluding with some open questions.

\section{Notation and Framework}\label{sec:frame}

In this section we formulate the basic framework and review the main
results from \cite{Rahn:02a}, which should be consulted for details.
Quantum states are represented by their density matrices.

The error correction process consists of three parts: \emph{encoding}
$\mathscr{Q}$, noise $\mathscr{N}$, and \emph{decoding} $\mathscr{D}$.
Each part is modeled as a \emph{quantum channel}, namely, a map taking
density matrices to density matrices. Quantum channels are required to
be linear, trace-preserving, and completely positive, hence of the
form
\begin{equation}\label{eq:channel}
\rho \rightarrow \sum_j A_j \rho A_j^\dagger, \ \ \ \ \ \ \ \ \
\mathrm{with} \quad \ \sum_j A_j^\dagger A_j = I,
\end{equation}
where $A_j$ are linear operators and $I$ is the identity
(cf.~\cite{Chuang:00}). In addition, we will assume that the channels are time-independent in order to simplify the study of their convergence. In the subsequent sections, we will often
denote quantum channels by $\$$.

Encoding $\mathscr{E}$ takes an initial logical qubit state $\rho_0$
to the initial register state $\rho(0)$ which evolves according to
some continuous-time noise dynamics. We consider the evolution for a
fixed amount of time $t$, turning noise into a discrete-time operation
$\mathscr{N}$ which takes $\rho(0)$ into a final register state
$\rho(t) = \mathscr{N}(\rho(0))$. Finally, decoding $\mathscr{D}$
takes $\rho(t)$ to the final logical qubit state $\rho_f$. The map
\begin{equation*}\label{eq:map}
\mathscr{G} = \mathscr{D} \circ \mathscr{N} \circ \mathscr{E}:
\rho_0 \rightarrow \rho_f
\end{equation*}
describes the effective dynamics of the encoded information resulting
from the physical dynamics of $\mathscr{N}$ and is called the
\emph{effective channel}.

We consider noise models $\mathscr{N}$
on $n$ qubits consisting of uncorrelated noise $\mathscr{N}^{(1)}$ on
each single physical qubit, so
$$
\mathscr{N} = \overbrace{\mathscr{N}^{(1)} \otimes \ldots \otimes
  \mathscr{N}^{(1)}}^{n \ \mathrm{times}}.
$$
Given an $n$ qubit quantum error correcting code $C$ with encoding
operation $\mathscr{E}$ and decoding operation $\mathscr{D}$, the map
taking the single qubit noise $\mathscr{N}^{(1)}$ to the effective
channel $\mathscr{G}$,
\be \label{eq:codingmap}
\Omega^C : \mathscr{N}^{(1)} \rightarrow \mathscr{D} \circ \left(
  \mathscr{N}^{(1)} \right)^{\otimes n} \circ \mathscr{E},
\ee
is called the \emph{coding map} of $C$.

The density matrix of one qubit can be expanded in the standard Pauli
basis $\mathscr{P} = \{I, X, Y, Z\}$ for density matrices and
represented as a four-dimensional real vector.  A noise channel
$\mathscr{N}^{(1)}$ can then be represented as a $4 \times 4$ matrix
\begin{equation}
\label{noise}
\mathscr{N}^{(1)}=
\bordermatrix{\cr
\cr
&1 & 0 & 0 & 0 \cr
&N_{XI} & N_{XX} & N_{XY} & N_{XZ} \cr
&N_{YI} & N_{YX} & N_{YY} & N_{YZ} \cr
&N_{ZI} & N_{ZX} & N_{ZY} & N_{ZZ} \cr
}.
\end{equation}
Zeroes in the first row are due to trace preservation. For an
arbitrary $n$ qubit code $C$, the entries of the matrix $\mathscr{G} =
\Omega^C (\mathscr{N}^{(1)})$ can be calculated to be
\begin{equation}\label{eq:Gmap}
\mathscr{G}_{\sigma \sigma'} = \sum_{\mu} \sum_{\nu}
\beta_\nu^\sigma \alpha_\mu^{\sigma'} \prod_{i=1}^n N_{\nu_i \mu_i},
\end{equation}
where $\mu = (\mu_1,\ldots,\mu_n)$, $\nu = (\nu_1,\ldots,\nu_n)$ run
over $\mathscr{P}^{\otimes n}$, and $\alpha_\mu^{\sigma'}$,
$\beta_\nu^\sigma$ are the coefficients in the expansions for the
encoding and decoding operations relative to $\mathscr{P}^{\otimes
  n}$. See \cite{Rahn:02a} for details.

If the matrix (\ref{noise}) is diagonal, $\mathscr{N}^{(1)}$ is called
a \emph{diagonal channel}. In that case, we write $x = N_{XX}$, $y =
N_{YY}$, and $z = N_{ZZ}$ and denote the channel by $[x,y,z]$. It was
shown in \cite{king:01} that complete positivity of such channels
implies that the point $(x,y,z)$ must be in the tetrahedron $\Delta$
defined by
\begin{equation}
  \label{eq:1}
\begin{split}
-x + y + z & \leq  1  \\
x - y + z & \leq  1  \\
x + y - z & \leq  1  \\
-x - y - z & \leq  1.
\end{split}
\end{equation}
It is easily checked that a \emph{single-bit Pauli channel} with
exclusive probabilities $0 \leq p_X, p_Y, p_Z \leq 1$,
$$
\rho \rightarrow (1 - p_X - p_Y - p_Z) \rho + p_X X \rho X + p_Y Y \rho
Y + p_Z Z \rho Z,
$$
has the following representation in the above notation:
$$
[1 - 2(p_Y + p_Z), 1 - 2(p_X + p_Z), 1 - 2(p_X + p_Y)].
$$
In fact, any diagonal channel can be realized as a single-bit Pauli
channel, so the parametrizations of $\Delta$ via $[x,y,z]$ and via
$(p_X, p_Y, p_Z)$ are equivalent.

The $n$ dimensional Pauli group is $\mathscr{P}_n=\{ \pm 1 , \pm i \}
\otimes \mathscr{P}^{\otimes n}$. Suppose we have a stabilizer code
that encodes $k$ qubits into $n$. Its stabilizer $S$ is an abelian
subgroup of $\mathscr{P}_n$ with $n-k$ generators $g_i$. The $2^k$
dimensional codespace is defined as
\begin{equation*}
C_S = \{ \ket{\psi} \in \left(\mathbb{C}^2\right)^{\otimes n} \text{ so that } g \ket{\psi} = \ket{\psi} \text{ for all }g\in S\}.
\end{equation*}
The subset of $\mathscr{P}_n$ that commutes with $S$ is the centralizer, and it includes encoded operations we can perform on the codespace. We measure each generator $g_i$, and let $\beta_i=0$ if we project into the $+1$ eigenspace, and $\beta_i=1$ if we project into the $-1$ eigenspace. We then have an error syndrome $\beta \in F_2^{n-k}$, and we correct with a recovery operator $R_{\beta} \in \mathscr{P}_n$.

It was shown in \cite{Rahn:02a} that if $C$ is a stabilizer code, then
$\Omega^C$ takes diagonal channels to diagonal channels. In fact, if
$S_1, \ldots, S_m$ are the generators of $C$, then
$$
\Omega^C [x,y,z] = \left[ \Omega_X^C (x,y,z), \Omega_Y^C (x,y,z),
  \Omega_Z^C (x,y,z) \right],
$$
where
$$
\Omega_\sigma^C [x,y,z] = \frac{1}{m} \sum_{k=1}^m f_{k \sigma}
x^{w_X(S_k \bar{\sigma})} y^{w_Y(S_k \bar{\sigma})} z^{w_Z(S_k
  \bar{\sigma})},
$$
\be \label{eq:homo}
f_{k \sigma} = \sum_j \eta(S_k,R_j) \eta(R_j, \bar{\sigma}),
\ee
and $\eta(\sigma, \sigma') = \pm 1$, if $\sigma \sigma' = \pm \sigma'
\sigma$, for $\sigma, \sigma' \in \{I, X, Y, Z \}$. Here, $w_\sigma$
denotes the $\sigma$-weight, $\bar{\sigma}$ is the encoded $\sigma$,
and the $R_j$ denote recovery operators corresponding to the error
syndromes. For later purposes, we extend $\eta$ as the natural
homomorphism to the negative of the Pauli matrices by $\eta(-\sigma,
\sigma') =
\eta(\sigma,-\sigma')=-\eta(\sigma,\sigma')=\eta(-\sigma,-\sigma')$.

Therefore, the components of $\Omega^C[x,y,z]$ are polynomials of
degree $n$ in $x,y,z$. Observe, however, that in general $\Omega^C$ is
a map from a higher dimensional space of non-diagonal channels to
itself. Non-diagonal channels of particular interest to us are
\emph{unital channels}; a channel $\mathscr{U}$ is unital if
$\mathscr{U}(I) = I$.

An important result from \cite{Rahn:02a} is that concatenation of
codes translates into composition of coding maps. In other words, if
$C_1$ and $C_2$ are codes and $C_1 \circ C_2$ denotes their
concatenation, then
$$
\Omega^C = \Omega^{C_1} \circ \Omega^{C_2}.
$$
Given a noise model $\mathscr{N}^{(1)}$ and code $C$, we are
interested in what this noise looks like under repeated concatenation
of the code $C$ with itself. Then the question is, does
\begin{equation*}
\label{gotoident}
{\Omega^C}^{\circ k}(\mathscr{N}^{(1)}) \to I, \quad \text{as $k \to \infty$?}
\end{equation*}
If this is the case, $C$ corrects the error given by
$\mathscr{N}^{(1)}$.

Rahn et al. \cite{Rahn:02a} focus mostly on the symmetric depolarizing
channel given in the above notation by $[e^{-\gamma t}, e^{-\gamma
  t},e^{-\gamma t}]$ and derive threshold estimates for various codes.
We take a global point of view, where instead of looking at noise
channels point by point, we consider the behavior of the coding map as
a discrete-time dynamical system and study the set of \emph{all} noise
channels attracted to the identity channel under iteration of the
coding map.  This approach enables us to use methods from the theory
of dynamical systems.

\section{Open set of correctable diagonal errors} \label{sec:diagonal}

We will first focus on diagonal noise channels, i.e., those given by a
diagonal matrix, as discussed in the previous section. The standing
assumption of this section is therefore that all noise channels are
diagonal. We saw that we can characterize the asymptotic properties of
the coding scheme involving the concatenation of a fixed code $C$ with
itself by studying the long-term behavior of the dynamical system
$$
\Omega^C : \Delta \rightarrow \Delta.
$$
We now review some necessary basics from the theory of dynamical
systems. Good introductory references are \cite{kh2003} and
\cite{pm82}.

\subsection{Dynamical systems preliminaries}

A (discrete-time) \emph{dynamical system} is a map $f: M \to M$, where
$M$ is a space with a certain additional structure (topological,
metric, differentiable, etc.). In our case, it suffices to assume that
$M$ is some Euclidean space $\R^k$ or a subset of it, and that $f$ is
a differentiable map.  We denote by $\mathbf{D}f(p)$ the derivative of
$f$ at a point $p$ and think of it as a linear operator on $\R^k$. We
will denote by $\| \mathbf{D}f(p) \|$ the norm of $\mathbf{D}f(p)$ as
such on operator; that is,
\begin{equation*}
\|\mathbf{D}f(p)\| = \max\{ \|\mathbf{D}f(p) v\|: \| v \| \leq 1\}.
\end{equation*}
(The norm on $\R^k$ is arbitrary but fixed.) If $\mathbf{D}f(p)$
depends differentiably on $p$, we define the second derivative of $f$
in the usual way as $\mathbf{D}^2 f = \mathbf{D}(\mathbf{D} f)$;
recall that $\mathbf{D}^2 f(p)$ can be thought of a bilinear map $\R^k
\times \R^k \to \R^k$ and $\| \mathbf{D}^2 f(p) \|$ then denotes its
norm.  Continuing recursively, we say that $f$ is of class $C^r$ (or
simply $C^r$) if $\mathbf{D}^r f(p)$ exists and is a continuous
function of $p$.

For $p \in M$, the set $\{ f^n(p): n
= 0,1,2,\ldots \}$, where $f^n = f \circ \cdots \circ f$ ($n$ times),
is called the \emph{orbit} or \emph{trajectory} of $f$. A fundamental
question in the theory of dynamical systems is: \textit{what is the
  long term behavior of trajectories}? That is, where does $f^n(p)$
end up eventually, as $n \to \infty$? The set of accumulation points
of the orbit of $p$ is called the $\omega$-limit set of $p$. An
example of such a set is a \emph{fixed point} of $f$, i.e., a point
$p$ such that $f(p) = p$.  A fixed point $p$ is \emph{locally
  attracting} if there exists a neighborhood $V$ of $p$ in $M$ such
that for every $x \in V$, $f^n(x) \to p$, as $n \to \infty$.  A basic
criterion for a fixed point to be locally attracting is the following.

\begin{lemma}
  Suppose $U \subset \R^k$ is open, $f: U \rightarrow \R^k$ is a $C^1$
  map, $p \in U$ is a fixed point of $f$, and $\lambda_0 = \|
  \mathbf{D}f(p) \| < 1$.  Then $p$ is locally attracting.
\end{lemma}
\begin{proof}
  Let $\lambda_0 < \lambda < 1$. Since $\mathbf{D}f(x)$ depends
  continuously on $x$ and $\| \mathbf{D}f(p) \| < 1$, there exists a
  neighborhood $V$ of $p$ in $U$ such that $\| \mathbf{D}f(x) \| \leq
  \lambda$, for all $x \in V$. Then, by the Mean Value Theorem,
\begin{equation*}
\| f(x) - f(p) \| \leq \lambda \| x - p \|,
\end{equation*}
for all $x \in V$. Therefore,
\begin{align*}
\|f^n(x) - p \| & = \| f^n(x) - f^n(p) \| \\
 & \leq \lambda^n \| x - p \| \\
 & \to 0,
\end{align*}
as $n \to \infty$.
\end{proof}
The largest such set $V$ is called the \emph{basin of attraction} of
the fixed point $p$, denoted by $\mathscr{B}(p)$. Let $B(x,r)$ denote
the \emph{open} ball of radius $r$ centered at $x$.
\begin{lemma}    \label{l:ball}
  Assume $f$ is $C^2$, the hypotheses of the previous lemma are
  satisfied, and $\| \mathbf{D}^2 f(x) \| \leq K$, for all $x \in U$.
  Then $B(p,(1-\lambda_0)/K) \cap U \subset \mathscr{B}(p)$.
\end{lemma}
\begin{proof}
  The proof goes along similar lines as the previous one. Let
  $\lambda_0 < \lambda < 1$ be arbitrary and $0 < r < (\lambda -
  \lambda_0)/K$. For an arbitrary point $x$ in the closed ball $B[p,r]
  \cap U$, we have
  \begin{eqnarray}
    \| \mathbf{D}f(x) \| & \leq & \| \mathbf{D}f(x) - \mathbf{D}f(p) \| + \| \mathbf{D}f(p) \| \nonumber \\
    & \leq & Kr + \lambda_0 \nonumber \\
    & \leq & \lambda, \nonumber
  \end{eqnarray}
  that is, $f$ is a contraction on $B[x,r] \cap U$. Furthermore, for
  all $x \in B[p,r] \cap U$,
\begin{eqnarray}
  \| f(x) - p \| & = & \| f(x) - f(p) \| \nonumber \\
  & \leq & \lambda \| x - p \| \nonumber \\
  & \leq r, \nonumber
\end{eqnarray}
which implies that $ B[p,r] \cap U$ is $f$-invariant. Therefore, under
iteration of $f$, every point in $B[p,r] \cap U$ converges to $p$, so
$B[p,r] \cap U \subset \mathscr{B}(p)$. Taking the union over all
$\lambda \in (\lambda_0, 1)$ proves the claim.
\end{proof}
Now take $f = \Omega^C$ and observe that $[1,1,1]$ is always an
isolated fixed point of $\Omega^C$, though not necessarily attracting.
For instance, $[1,1,1]$ is a saddle for the coding map
$\Omega^\mathrm{bf}$ of the bit-flip code. However, if $C$ is the Shor
or five-bit code, then $\mathbf{D} \Omega^C [1,1,1] = \mathbf{0}$, so
$[1,1,1]$ is locally attracting. The following result shows that this
is not a coincidence.

\begin{proposition}
\label{prop:zero}
Under the assumptions above, if $C$ is a quantum error correcting code
of distance $\geq 3$, then
$$
\mathbf{D}\Omega^C [1,1,1] = \mathbf{0}.
$$
\end{proposition}
\begin{proof}
  It suffices to show that $\mathbf{D}\Omega^C$ sends three linearly
  independent vectors to zero.

  Since the distance of the code is at least three, $C$ corrects all
  errors of weight one. In particular, it corrects all single-bit
  Pauli channel errors
  $$
  \rho \rightarrow (1-\epsilon) \rho + \epsilon \sigma \rho \sigma,
  $$
  for $\sigma \in \{X, Y, Z \}$ and $0 \leq \epsilon \leq 1$. Such errors
  correspond to noise channels $[1, 1-2\epsilon, 1-2\epsilon]$, $[1-2\epsilon, 1, 1-2\epsilon]$,
  and $[1-2\epsilon, 1-2\epsilon, 1]$, for $\sigma = X, Y, Z$, respectively. Let us
  consider $\sigma = X$. To say that $C$ corrects $X$-errors means
  that
$$
\Omega^C [1, 1-2\epsilon, 1-2\epsilon] = [1, 1-O(\epsilon^2), 1-O(\epsilon^2)].
$$
This implies that the directional derivative
$$
\mathbf{D}\Omega^C [1,1,1] v_X = \left. \frac{d}{d\epsilon}
\right|_{\epsilon=0} \Omega^C ([1,1,1] + \epsilon v_X) = 0,
$$
where $v_X = (0, -1, -1)^T$. Similarly, we can show that $\mathbf{D}\Omega^C
[1,1,1] v_Y = \mathbf{D}\Omega^C [1,1,1] v_Z = 0$, where $v_Y = (-1,0,-1)^T$
and $v_Z = (-1,-1,0)^T$. Since $v_X, v_Y, v_Z$ are linearly
independent, it follows that $\mathbf{D}\Omega^C [1,1,1] = \mathbf{0}$.
\end{proof}

\begin{corollary}
  For every code $C$ of distance at least three, $[1,1,1]$ is an
  attracting fixed point of the coding map $\Omega^C: \Delta \to
  \Delta$. If $\mathscr{B}_C$ denotes its basin of attraction and $\|
  \mathbf{D}^2 \Omega^C \| \leq K$ on $\Delta$, then
\begin{equation}   \label{cor:basin}
B\left([1,1,1],\frac{1}{K}\right) \cap \Delta \subset \mathscr{B}_C.
\end{equation}
\end{corollary}
\begin{proof}
  Observe that $\Omega^C$ can be extended to the whole space $\R^3$,
  has $[1,1,1]$ as a fixed point, and, by Proposition~\ref{prop:zero},
  $\lambda_0 = \mathbf{D}\Omega^C[1,1,1] = \mathbf{0}$. Therefore,
  $[1,1,1]$ is locally attracting for $\Omega^C$ as a map $\R^3 \to
  \R^3$. By Lemma~\ref{l:ball}, $B([1,1,1],1/K)$ is contained in the
  basin of attraction of $[1,1,1]$, again as a fixed point of
  $\Omega^C :\R^3 \to \R^3$. However, we know that $\Delta$ is an
  \emph{invariant set} for $\Omega^C$, i.e., $\Omega^C (\Delta)
  \subset \Delta$, and it contains $[1,1,1]$. Therefore, points in
  $B([1,1,1],1/K) \cap \Delta$ are both attracted to $[1,1,1]$
  \emph{and} stay in $\Delta$ under iteration of $\Omega^C$. This
  proves \eqref{cor:basin}.
\end{proof}

\begin{proposition}

  Suppose $C$ is a CSS code. It will be shown in
  Theorem~\ref{cssfuncs} that
  $$
  \Omega^C [x,y,z] = [f(x), g(x,y,z), f(z)],
  $$
  for some polynomials $f, g$. Let $a$ be the largest fixed point of
  $f$ in $(0,1)$. Then
  $$
  \mathscr{B}_C = \{ [x,y,z] \in \Delta : x > a, \ z > a \}.
  $$
\end{proposition}
\begin{proof}
  It follows from Proposition~\ref{prop:zero} that $1$ is an
  attracting fixed point of $f$. Let $(\alpha,\beta)$ be its basin of
  attraction. It is well known that its boundary $\{\alpha, \beta\}$
  is $f$-invariant. Since $\alpha \in [a,1)$ and $[a,1)$ is
  $f$-invariant, it follows that $\alpha$ is a fixed point of $f$.
  Therefore, $\alpha = a$. This means that for every $x \in (a,1)$,
  $f^k(x) \rightarrow 1$, as $k \rightarrow \infty$.

  Now suppose $[x,y,z] \in \Delta$, $x > a, z > a$. Then
  $$
  \left(\Omega^C\right)^k [x,y,z] = [f^k(x), y_k, f^k(z)].
  $$
  We know that $f^k(x), f^k(z) \rightarrow 1$. Let $y_*$ be an
  accumulation point of the sequence $(y_k)$. Since $[1,y_*,1] \in
  \Delta$, it follows that $y_* = 1$. Therefore,
  $\left(\Omega^C\right)^k [x,y,z] \rightarrow [1,1,1]$, as $k
  \rightarrow \infty$, which implies $\{ [x,y,z] \in \Delta : x > a, \
  z > a \} \subseteq \mathscr{B}_C$.

  To show the opposite inclusion, assume the contrary, i.e., that
  there exists a point $p = [x,y,z] \in \mathscr{B}_C$ such that $p
  \not \in \{ [x,y,z] \in \Delta : x > a, \ z > a \}$. Then $x \leq a$
  or $z \leq a$. In the former case, $f^k(x)$ does not converge to 1,
  and in the latter, $f^k(z) \not \rightarrow 1$, contrary to our
  assumption that $p$ is in the basin of attraction of $[1,1,1]$.
\end{proof}

\section{Faulty gates}\label{sec:faulty}

We want to extend the analysis in \cite{Rahn:02a} to include faulty
gate operations both in the error correction and in the computation
circuits. Gate errors are a common form of noise in quantum information processing. We show how to incorporate faulty gates into the current framework and how they change the effective channel and the coding map. Note that fault
tolerance for our noise model has been shown, but that there is some dispute
about the validity of that model and whether quantum fault-tolerance is
possible \cite{Alicki}.

\subsection{A simple noise model}\label{sec:simple}

Our first approach is to start with a very simple error
model for faulty unitary gates $G$:
\begin{equation}\label{eq:faultgate}
G: \rho \longrightarrow (1-\epsilon)G \rho G^\dagger + \epsilon \frac{1}{N} I.
\end{equation}
This error model is rather generic. It has the additional advantage that noise from sequential gates is {\em additive}; if we combine two faulty operations as in Eq. (\ref{eq:faultgate}), we obtain
\begin{eqnarray}\label{eq:add}
G_2 \circ G_1: \rho &\longrightarrow & G_2\left((1-\epsilon_1)G_1 \rho G_1^\dagger + \frac{\epsilon_1}{N} I\right)\nonumber\\
&=&(1-\epsilon_2)(1-\epsilon_1)G_2G_1 \rho G_1^\dagger G_2^\dagger \nonumber\\&+&(1-\epsilon_2)\frac{\epsilon_1}{N} I+\frac{\epsilon_2}{N}I \nonumber\\
&\approx& (1-\epsilon_1-\epsilon_2) G_2G_1 \rho (G_2 G_1)^\dagger \nonumber\\&+& \frac{\epsilon_1+\epsilon_2}{N}I,
\end{eqnarray}
i.e. a faulty process with $\epsilon=\epsilon_1+\epsilon_2$.
As we have seen, the effective dynamics of one level of concatenation is simply encoding, noise and decoding, i.e.
$${\cal G}={\cal D} \circ {\cal N} \circ {\cal E}.$$

Let us also assume here that the noise on the qubits is unital, i.e. ${\cal N}(I)=I$.
We now show that faulty gates in this model have the same effect as noise; hence we can effectively treat noise from faulty gates and other types of noise on the qubits in the same way.

The encoding operation can be written concisely as
${\cal E}(\rho)=B \rho B^\dagger$, where $B=\ket{\bar 0}\bra{0}+\ket{\bar 1}\bra{1}$ (or, for codes that encode more than one qubit, $B=\sum_i \ket{\bar i}\bra{i}$). This encoding is performed by applying a sequence of gates, possibly faulty, as in Eq. (\ref{eq:faultgate}). The operation corresponding to $B$ can be implemented with unitary gates in a larger space by appending some ancillary qubits, for instance as $U_B: \ket{i}\ket{0} \longrightarrow \ket{\bar i}$. If errors occur according to Eq. (\ref{eq:faultgate}), the resulting operation will be ${\cal E}_{\epsilon_E} : \rho \rightarrow (1-\epsilon_E) U_B \rho U_B^\dagger + \frac{\epsilon_E}{N} I=(1-\epsilon_E){\cal E}(\rho)+\frac{\epsilon_E}{N}I$, where $\cal E$ denotes the error-free encoding and $\epsilon_E$ is the noise accumulated from gates during encoding. In an analogous way it can be seen that a decoding map $\cal D$, implemented with faulty gates, can be written as ${\cal D}_{\eps_D} : \rho \rightarrow (1-\epsilon_D) {\cal D}(\rho) +\frac{\epsilon_D}{2}I$, where we have used that ${\cal D}: \frac{1}{N}I \longrightarrow \frac{1}{2} I$. Putting this together under the simplifying assumption that ${\cal N}(I)=I$ (unital channels), and using additivity of error from faulty gates, we get
$$\rho \longrightarrow (1-\epsilon) {\cal
  G}(\rho)+\frac{\epsilon}{2}I,$$
where $\epsilon=\epsilon_D +
\epsilon_E$ and ${\cal G}$ is the effective channel with perfect
gates. In other words, faulty gates only contract the iterated map by
$(1-\epsilon)$. As a result, the coding map $\Omega^C$ (see Eq.
(\ref{eq:codingmap})) changes to $\Omega^C_f$, the coding map with
faulty gates, as
$$
\Omega_f^C: {\cal N} \longrightarrow (1-\eps) {\cal D} \circ {\cal N} \circ {\cal E} +\eps \frac{1}{2}I=(1-\eps)\Omega_C +\eps  \frac{1}{2}I.
$$
The entries of the matrix ${\cal G}$ for the coding map change as
\begin{equation}  \label{faulty-map}
{\cal G}^f_{\sigma \sigma'}=(1-\eps) G_{\sigma
  \sigma'}+\frac{\eps}{2} \delta_{\sigma 1} \delta_{\sigma' 1},
\end{equation}
where we have used the fact that the coding map whose only non-zero
entry is $G_{11}$ represents a mapping of $\rho$ to the identity
matrix. In other words, the incorporation of faulty gates into our
analysis results in an affine mapping of the coding map: $G$ is
contracted by $(1-\eps)$ and the element $\eps \delta_{11}$ is added.

\subsection{More general noise}

It is not difficult to extend this analysis to more general noise in the gates and general noise on the qubits. Let us assume that instead of the restricted noise model of Eq. (\ref{eq:faultgate}) we are dealing with generic noise of rate $\eps$. We can write
\begin{equation*}
G: \rho \longrightarrow (1-\epsilon)G \rho G^\dagger + \epsilon N_G(\rho),
\end{equation*}
where $N_G$ is some general noise operation.

The analysis of the previous Section \ref{sec:simple} goes through line by line. The noise process is additive (with  $I/N$ in Eq. (\ref{eq:add}) replaced by $\eps_1 G_2 N_{G_1}(\rho) G_2^\dagger +\eps_2 N_{G_2}(\rho)$). The encoding and decoding operations can then be written as
\begin{eqnarray*}
{\cal E}_{\epsilon_E} &:& \rho \rightarrow (1-\epsilon_E) U_B \rho U_B^\dagger + \frac{\epsilon_E}{N} I\\&=&(1-\epsilon_E){\cal E}(\rho)+\epsilon_E N_E(\rho)\nonumber \\
{\cal D}_{\eps_D} &:& \rho \rightarrow (1-\epsilon_D) {\cal D}(\rho) +\epsilon_D N_D(\rho),
\end{eqnarray*}
where $N_E$ and $N_D$ are the noise resulting from encoding resp. decoding. Concatenating yields
$$
\rho \longrightarrow (1-\epsilon) {\cal G}(\rho)+\epsilon N_{DE}
$$
with $\eps = \eps_E + \eps_D$ and the cumulative noise can be written to first order as
$$\eps N_{DE}=\eps_E {\cal D}( {\cal N}(N_{E}(\rho)))  +\eps_D N_{D}({\cal N}({\cal E}(\rho))$$.
The new coding map with faulty gates is then very similar to before:
$$
\Omega_f^C: {\cal N} \longrightarrow (1-\eps) {\cal D} \circ {\cal
  N} \circ {\cal E} +\eps N_{DE}(\rho)=(1-\eps)\Omega_C +\eps
N_{DE}.
$$
In other words, faulty gates introduce a perturbation to the
original coding map studied in the previous section. They can be
treated in the same way as noise on the qubits. In fact we see that
the occurrence of faulty gates is the same as a process with increased
noise on the gates and perfect gates. However, if the noise on gates
is small compared to the noise on qubits, we can treat it as a
perturbation to the original coding map. We will show how to
incorporate such perturbations in the analysis with the following
Lemma. Here, $\|h\|_{C^1}$ denotes the $C^1$ norm of a smooth map $h$
on its domain, that is, the maximum of the suprema of $|h|$ and $\| \mathbf{D}h
\|$.

\begin{lemma}  \label{l:perturb}

  Suppose $U \subset \R^n$ is an open set, $f:U \rightarrow \R^n$ is
  smooth (at least $C^2$), $f(p) = p$ and $\lambda = \| \mathbf{D}f(p) \| < 1$.
  Then for small enough $\epsilon > 0$ and every smooth map $g: U
  \rightarrow \R^n$, if $\| g - f \|_{C^1} < \epsilon$, then $g$ has a
  fixed point $q$ such that $\| \mathbf{D}g(q) \| < 1$ and $|q-p| < \epsilon/(1
  - \lambda)$.

  In other words, if a map has an attracting fixed point, then any
  sufficiently small $C^1$ perturbation of it also has an attracting
  fixed point which is close to the original one.

\end{lemma}
This is a standard fact from the theory of dynamical systems; for
completeness, we supply a proof here.
\begin{proof}
  Let $M$ be an upper bound of $\| \mathbf{D}^2 f \|$ on some relatively
  compact neighborhood $V$ of $p$. Since $\lambda < 1$, there exists
  $r > 0$ such that $f$ maps the closed ball $B[p,r]$ into itself and
  $B[p,r] \subset V$. Without loss, we can take $r$ so small that $r <
  (1-\lambda)/M$. Assume $0 < \epsilon < \min((1-\lambda) r, 1 -
  \lambda - Mr)$. Then it is not difficult to show that for every $x
  \in B[p,r]$, $|g(x) - p| \leq \epsilon + \lambda r < r$, which means
  that $g$ takes $B[p,r]$ into itself.  Therefore, by the Brouwer
  fixed point theorem, $g$ has a fixed point, say $q$, in $B[p,r]$.
  Since
  \begin{eqnarray}
    |q - p| & = & |g(q) - p| \nonumber \\
            & \leq & |g(q) - f(q)| + |f(q) - p| \nonumber \\
            & \leq & \epsilon + \lambda |q - p|, \nonumber
  \end{eqnarray}
we obtain $|q-p| < \epsilon/(1-\lambda)$.

To show that $q$ is an attracting fixed point for $g$, let us show
that $\| \mathbf{D}g(q) \| < 1$. Observe first that $\| \mathbf{D}f(q) \| \leq Mr +
\lambda < 1 - \epsilon$. Therefore, $\| \mathbf{D}g(q) \| \leq \| \mathbf{D}g(q) - \mathbf{D}
f(q) \| + \| \mathbf{D} f(q) \| < 1$.
\end{proof}
It is clear from (\ref{faulty-map}) that the coding map $\Omega^C_f$
of a code with faulty gates is a $C^1$ small perturbation of the
coding map $\Omega^C$ with perfect gates.

\section{Analysis of Channels}\label{sec:channels}

In this section we will give several technical results about channel maps, which we will subsequently use to analyze various diagonal and non-diagonal channels and to give examples. In particular we will study in detail how non-diagonal elements of a noise channel affect its convergence and threshold.

\subsection{The two-point theorem}

We look at bounds for a general channel, resulting in Thm. \ref{altconv}.

\begin{lemma}
For any non-identity Pauli matrix $\sigma$,
\begin{eqnarray}
\label{rowres}
N_{\sigma X}^2 + N_{\sigma Y}^2 + N_{\sigma Z}^2 \leq (1-|N_{\sigma I}|)^2\\
\label{colres}
(N_{XI} \pm N_{X \sigma})^2 + (N_{YI} \pm N_{Y \sigma})^2 \\+ (N_{ZI} \pm  N_{Z \sigma})^2 \leq 1.\nonumber
\end{eqnarray}
All elements of the channel are real.
\end{lemma}
\begin{proof}
$\mathscr{N}$ preserves hermiticity, and is positive (sends non-negative $\rho$ to non-negative $\rho$) \cite{Pres}. The first condition implies that the elements are real. Then the adjoint channel, which has the map $\mathscr{N^{\dagger}}\rho  =  \sum_k A_k^{\dagger} \rho  A_k\nonumber$, is also positive. A simple calculation shows that a matrix $\rho = c_I I + c_X X + c_Y Y + c_Z Z$ is non-negative if and only if $c_I \geq \sqrt{c_X^2 +c_Y^2 +c_Z^2}$.

Let $c = \sqrt{N_{\sigma X}^2 + N_{\sigma Y}^2 + N_{\sigma Z}^2}$, and apply
\begin{equation*}
\mathscr{N} (c I \pm (N_{\sigma X}X + N_{\sigma Y}Y + N_{\sigma Z}Z)),
\end{equation*}
which gives $c_I = c$, and $c_\sigma = c N_{\sigma I} \pm c^2$, so the non-negative condition gives $\abs{c N_{\sigma I } \pm c^2} \leq c$, from which we get $c^2 \leq (1-\abs{N_{\sigma I}})^2$, which gives equation \ref{rowres}.

Let $b_{\sigma \sigma'}= N_{X \sigma} N_{X \sigma'} + N_{Y \sigma} N_{Y \sigma'} + N_{Z \sigma} N_{Z \sigma'}$. Now let $c=\sqrt{b_{II} + b_{\sigma \sigma} \pm 2 b_{I \sigma}}$. Then, apply $\mathscr{N^{\dagger}}$ to
\begin{equation*}
c I - (N_{XI} X + N_{YI} Y + N_{ZI}Z) \pm (N_{X \sigma} X + N_{Y \sigma}Y + N_{Z \sigma}Z),
\end{equation*}
which gives $c_I = c - b_{II} \pm b_{I \sigma}$ and $c_{\sigma}= -b_{I \sigma} \pm b_{\sigma \sigma}$, so $c-b_{II} \pm b_{I \sigma} \geq \abs{-b_{I \sigma} \pm b_{\sigma \sigma}}$, which gives $c \geq b_{II} + b_{\sigma \sigma} \pm  2 b_{I \sigma}=c^2$, so $c \leq 1$,  which gives equation \ref{colres}.

This proof extends naturally to multi-qubit channels.

\end{proof}
\begin{corollary}
\label{rownorm}
Each row of a quantum channel $\mathscr{N}$ in the Pauli basis has norm at most $1$.
\end{corollary}
\begin{proof}
Since $|N_{\sigma I}| \leq 1$, we have $1 - N_{\sigma I}^2 \geq (1-|N_{\sigma I}|)^2$, and so the result follows from Eq. (\ref{rowres}).
\end{proof}
\begin{corollary}
\label{columnnorm}
Let $A = N_{XI}^2 +N_{YI}^2 + N_{ZI}^2$ be the non-unital portion of the channel. Then we have that any other column of the channel in the Pauli basis has $L_2$ norm squared  $N_{X\sigma}^2 +N_{Y\sigma}^2 + N_{Z\sigma}^2 \leq 1-A$.
\end{corollary}
\begin{proof}
Follows immediately from Eq. (\ref{colres}).
\end{proof}

\begin{theorem}[Two-point theorem]
\label{altconv}
If two of $N_{XX}$, $N_{YY}$, $N_{ZZ}$ are $1$, then the channel is the identity channel.
\end{theorem}
\begin{proof}
Let $\sigma_1$, $\sigma_2$, $\sigma_3$ be some permutation of the Pauli matrices such that $N_{\sigma_1 \sigma_1}=N_{\sigma_2 \sigma_2}=1$.
From Corollary \ref{rownorm}, $N_{\sigma_1 \sigma_1}$ and $N_{\sigma_2 \sigma_2}$ are the only non-zero elements in their rows.
From Corollary \ref{columnnorm}, the non-unital part must be $0$, and $N_{\sigma_1 \sigma_1}$ and $N_{\sigma_2 \sigma_2}$ are the only non-zero elements in their columns.
It then follows that the channel is diagonal.
From the conditions on diagonal channels given in Eq. (\ref{eq:1}), it easily follows that if two terms are equal to $1$, the $3^{rd}$ term must equal $1$, and so we have the identity channel.
\end{proof}

\subsection{Example: Generalized Shor codes}

In this section we give give a first application of our formalism and the general bounds we obtained. We study generalized Shor codes, which are  bit flip and phase flip codes concatenated with each other. We will assume a diagonal channel $[x,y,z]$ in what follows. Note that Thm. \ref{altconv} is easy to prove in this case; it follows immediately from Eq. (\ref{eq:1}).

\paragraph{Bit flip, phase flip}

The $n$ qubit bit flip code is a classical code on $n$ qubits that corrects all bit flip errors on less than $\frac{n}{2}$ qubits and none of the errors on greater than $\frac{n}{2}$ qubits; if $n$ is even it also corrects half of the errors on exactly $\frac{n}{2}$ qubits. The coding map is $\Omega^{bf_n}[x,y,z]=[x^n,h_n(x,y,z),f_n(z)]$. To see this note that the code does not correct phase flips ($Y$ or $Z$ errors), and so if $p=p_Y+p_Z$, the $p$-component of the coding map must be a function of only $p$. Since $x=1-2(p_Y+p_Z)=1-2p$, it follows that the $x$-component of the coding map must be a function of only $x$. The only such element of the $X$ equivalence class gives us $x^n$.

To see that the $z$-component depends on $z$ only, note that the code can correct bit flips ($X$ or $Y$ errors), sending them to $I$ or $Z$ errors, respectively, and so if $p'=p_X+p_Y$, by similar reasoning as above we observe that the $p'$ component depends only on $p'$ and hence that the $z$-component is a function of only $z$. Now, assume only $X$ errors. Then $z=1-2p_X$, and $f_n(z)=1-2g(\frac{1-z}{2})$, where $g(p)$ is the failure probability as a function of an $X$ error rate of $p$. We can obtain $g(p)$ from the properties of the classical bit flip code.

 Since the function $h_n(x,y,z)$ does not affect the $x$ and $z$ components of the channel, from Thm. \ref{altconv}, we may ignore it for the purposes of convergence to the identity channel.

Some values of $f_n$ are
\begin{eqnarray*}
f_1(x)=f_2(x)=&x \cr
f_3(x)=f_4(x)=&\frac{3}{2}x-\frac{1}{2}x^3 \cr
f_5(x)=f_6(x)=&\frac{15}{8}x-\frac{5}{4}x^3+\frac{3}{8}x^5.
\end{eqnarray*}
For the phase flip code we get similarly $\Omega^{pf_n}[x,y,z]=[f_n(x),h'_n(x,y,z),z^n]$ by exchanging the roles of $x$ and $z$.

These codes will have two critical values, $x_c$ and $z_c$. If $x>x_c$ then $x \rightarrow 1$, and similarly for $z$.

\paragraph{Specific codes}
We can now obtain sharper results for the error threshold of concatenated bit flip and phase flip codes, extending \cite{Rahn:02a}.

The often discussed $[[9,1,3]]$ Shor code has the coding map: $\Omega^{Shor}[x,y,z]=\Omega^{pf_3} \Omega^{bf_3} [x,y,z]=[f_3^3(x),h''(x,y,z),f_3(z^3)]$. We define a $[[25,1,5]]$ code to be $\Omega^{25}=\Omega^{pf_5} \Omega^{bf_5}$, and a $[[15,1,3]]$ code to be $\Omega^{15}=\Omega^{pf_5} \Omega^{bf_3}$

The $[[25,1,5]]$ code has critical values of $x_c=0.916208$, and $z_c=0.645611$. The $[[15,1,3]]$ code has critical values of $x_c=0.794438$ and $z_c=0.850432$. If $x=z$, the $[[15,1,3]]$ code performs much better than the $[[25,1,5]]$, even though it is less redundant.

\subsection{Convergence of non-diagonal channels}
\label{sec:nondiagonal}
\label{coffbounds}
\label{nondiag}
\label{another}
In this section we will establish some general results for non-diagonal channels in the case of stabilizer codes \cite{Gottesman:97b}. Non-diagonal channels are in general much harder to analyze than their diagonal counterparts, as the parameters span a $12$-dimensional manifold. However, we will show that in certain cases these channels converge to diagonal channels, and will discuss when these converge to the identity channel.

We can decompose the single qubit noise operator $\mathscr{N}$ as
\begin{equation}
\label{almostdiag}
\mathscr{N}=L+\epsilon M,
\end{equation}
where $L$ is the diagonal part, and $\epsilon$ is chosen such that $M$ has no term with absolute value more than $1$; it contains the off-diagonal terms. We show that if $\epsilon$ is sufficiently small and $d \geq 3$, then repeated application of the coding map yields a diagonal matrix.
This will allow  to restrict our analysis to diagonal channels, at least in certain regimes.

We wish to analyze the absolute values of the difference that the non-diagonal terms make on the channel after we apply the coding map. Define the difference matrix
\begin{equation*}
\Gamma =\Omega^C(\mathscr{N})-\Omega^C(L).
\end{equation*}
Let us assume that the code is an $[[n,k,d]]$ stabilizer code \cite{Gottesman:97b} (it encodes $k$ qubits into $n$ qubits, and has distance $d$, which is the minimal weight of an undetected error). Let $m$ be the minimal weight of a non-identity stabilizer element.

\begin{theorem}
\label{nondiagterm}
The non-diagonal terms of the difference matrix $\Gamma$ have absolute value at  most $c_d \epsilon^d$.
The diagonal terms of  $\Gamma$ are at most $c_m \epsilon^m$ in absolute value.
These coefficients are bounded above by
\begin{equation}
\label{coffboundseq}
\max(c_d,c_m) \leq  2^{n-k} \sum_{\sigma''}|\mathscr{D}_{\sigma' \sigma''}| \leq 4^{n-k}.
\end{equation}
\end{theorem}
\begin{proof}
We can rewrite Eq. (\ref{eq:Gmap}) as
\begin{equation}\label{eq:Gmap2}
G_{\sigma \sigma'}={\cal D}_\sigma {\cal N} {\cal E}_{\sigma'},
\end{equation}
where $\mathscr{E}_\sigma$ is the $\sigma$ column of $\mathscr{E}$ and similarly for ${\cal D}$. The (non-zero) entries of $\mathscr{E}_I$ are the stabilizer elements, and the non-zero elements of $\mathscr{E}_\sigma$ are $\overline{\sigma}$ times the stabilizer elements, where  $\overline{\sigma}$ is the encoded $\sigma$.
We note that $\mathscr{E}_{\sigma' \sigma}$ is non-zero only if $\sigma'$ and $\overline{\sigma}$ are in the same equivalence class of $C(S)$ modulo $S$, where $S$ is the stabilizer group, and $C(S)$ is its centralizer (see \cite{Gottesman:97b} for more detailed definitions).

 Now the non-diagonal elements of $\Gamma$ depend on
  the non-zero elements of $\mathscr{E}_{\sigma}$ and
  $\mathscr{E}_{\sigma'}$ with $\sigma \neq \sigma'$, which correspond
  to the $\sigma$ and $\sigma'$ equivalence classes of $C(S)$,
  which differ on at least $d$ qubits. Then from Eq. (\ref{eq:Gmap}) resp. Eq. (\ref{eq:Gmap2}), it follows that the non-diagonal terms involve at least $d$ non-diagonal terms of ${\cal N}$ and are hence
  $O(\epsilon^d)$ from Eq. (\ref{almostdiag}).  The difference of the diagonal elements corresponds
  to elements of the same $\mathscr{E}_{\sigma}$, which differ on at least $m$
  qubits, since $m$ is the minimal weight of different
  elements in the same equivalence class (non-zero elements of the same
  $\mathscr{E}_{\sigma}$). Hence they are $O(\epsilon^m)$.

From Eq. (\ref{eq:Gmap2}) it is easy to see that the coefficients $c_d$ and $c_m$ are bounded above by
$$\sum_{\sigma'', \sigma'''} |\mathscr{D}_{\sigma \sigma''} \mathscr{E}_{\sigma''' \sigma'}| \leq \sum_{\sigma''} |\mathscr{D}_{\sigma \sigma''}| \sum_{\sigma'''}|\mathscr{E}_{\sigma''' \sigma'}| \leq 4^{n-k}, $$
where we used that each coefficient is at most $1$ in absolute value and the cardinality of the stabilizer group.
\end{proof}
Note that in certain cases we have explicit expressions for $\sum_{\sigma''}|\mathscr{D}_{\sigma' \sigma''}|$, which can come from calculations with a diagonal noise channel and can give us tighter bounds on $c_d$ and $c_m$ than the generic $4^{n-k}$.

\ignore{
If $\mathscr{N}$ is diagonal, then this simplifies to
\begin{equation}
\label{Gxy}
\mathscr{G}_{\sigma \sigma}=\sum_{\sigma' = \pm \overline{\sigma} \textrm{ modulo }S} \eta(Rav,\sigma') \mathscr{N}_{\sigma' \sigma'},
\end{equation}
where $S$ is the stabilizer group.

Now  $\sum_{\sigma''}|\mathscr{D}_{\sigma' \sigma''}|$ is the sum of the absolute values of the coefficients of a diagonal channel's coding map (as before we make the assumption that the channel is the same on each qubit), and so doesn't require much additional work to calculate.
From Eq. (\ref{coffboundseq}), this allows one to easily get the tighter upper bound of $c \leq 2^{n-k} \sum_{\sigma''}|\mathscr{D}_{\sigma' \sigma''}| $.

In general, this general bound still is significantly worse than if we calculated the coefficients directly from the general non-diagonal map. If some of the $12$ variables $N_{\sigma \sigma'}$ of the channel are $0$ (for example, if we have a unital channel), then the values of $c_t$ and $c_d$ becomes even smaller.
}

\paragraph{Convergence to the identity}
Suppose we concatenate the above coding map $i$ times. Then the absolute values of the off-diagonal terms are bounded above by $a_i$, where $a_0=\epsilon$, and $a_{n+1}=c_d a_n^d$. Then, from Thm. \ref{nondiagterm},
\begin{equation*}
a_i=c_d^{\sum_{j=0}^{i-1}d^j } \epsilon^{d^i}=\epsilon_0(\frac{\epsilon}{\epsilon_0})^{d^i},
\end{equation*}
where $\epsilon_0 = \sqrt[d-1]{\frac{1}{c_d}}$ is defined for $d > 1$.
Since these affect the diagonal terms by at most $c_m \epsilon^m$, we can bound the correction for the diagonal terms as
\begin{equation}
b_i=c_m a_{i-1}^m=c_mc_d^{t\sum_{j=0}^{i-2}d^j } \epsilon^{md^{i-1}}=c_m \epsilon_0^m (\frac{\epsilon}{\epsilon_0})^{md^{i-1}}.
\end{equation}
Now we assume that the non-diagonal terms go to $0$, which means that  $\epsilon < \epsilon_0$, and so $a_i$ and $b_i$ both go monotonically to $0$.
From Thm. \ref{nondiagterm}, we can see that if the map $\Omega^C(L^{\otimes n})-c_m\epsilon^m I$ converges to within $O(\epsilon^m)$ of the identity matrix, then so does $\Omega^C(\mathscr{N}^{\otimes n})$. However, we can get a tighter bound than this.

Let $L_0=[x_0,y_0,z_0]$ be the diagonal part of the channel. We define $L_i=\Omega^C(L_{i-1})-b_i I$. We can think of the $L_i$ as a lower bound on the diagonal part of the channel. Then, the channel goes to $[1,1,1]$, if $L_i \rightarrow [1,1,1]$. These coding maps are $\Omega^C_i(L)=\Omega^C(L_{i-1})-b_iI$, and $\Omega^C_1(L)=\Omega^C(L_0)-c_m \epsilon^m I$. The channel converges to identity if
\begin{equation*}
\ldots \circ \Omega^C_2 \circ \Omega^C_1 L = [1,1,1].
\end{equation*}

\subsection{CSS codes on $1$ qubit with a generalized noise channel}
\label{csscodes}
In this section we tighten our result  in the case of CSS codes \cite{Shor:95a,Steane:96a,Calderbank:96a}.

Let our code be a $[[n,1,d]]$ CSS code. From the construction of CSS codes from classical codes, $n$ must be odd. Its stabilizer group is generated by $n-1$ generators, half of which depend only on tensor products of $I$s and $X$s, and the other half are the same, except they have $Z$s replacing the $X$s. We can write the stabilizer group $S$ as the span of $\{S(X),S(Z)\}$, where $S(A) \in A_S$, and $A_S$ is the $n$-dimensional Pauli Matrices $\mathscr{P}_n$ which only depend on tensor products of $I$ and $A$. The stabilizer elements in $S(X)$ are used to correct against $Z$ errors, and the stabilizer elements of $S(Z)$ are used to correct against $X$ errors, and so we can write the set of recovery operators  as $R(\varepsilon_X,Z)$ and $R(\varepsilon_Z,X)$,
where $\varepsilon_A$ are the components of the syndromes obtained by measuring stabilizer generators from $S(A)$, and each $R(\varepsilon,A) \in A_S$.

The Pauli operators are encoded as
\begin{eqnarray}
\label{encodedpauli}
\overline{X}=X^{\otimes n} \in X_S\\
\overline{Z}=Z^{\otimes n} \in Z_S\nonumber\\
\overline{Y}=i\overline{X} \overline{Z}=(-1)^{\frac{n-1}{2}}Y^{\otimes n}\in Y_S. \nonumber
\end{eqnarray}
To obtain a convenient representation of the decoding operator ${\cal D}$, we
define the average recovery function as
\begin{equation*}
Rav= \frac{1}{|R_i|} \sum_i R_i,
\end {equation*}
where the $R_i$ are the recovery operators (see Sec. \ref{sec:frame}).
Let $\mathscr{T} \in \mathscr{M}_{2^n, 2^n}$ be the the diagonal matrix given by%
\begin{equation}\label{eq:recover}
\mathscr{T}_{\sigma \sigma} = \eta(Rav, \sigma)
\end{equation}
Where $\eta$ is the linear homomorphism defined in Sec. \ref{sec:frame} Eq. \ref{eq:homo}. In particular note that if $\sigma$ commutes with all recovery operators $R_i$, then $\mathscr{T}_{\sigma \sigma}=1$, and if $\sigma$ anti-commutes with all of the recovery operators then $\mathscr{T}_{\sigma \sigma}=-1$.
Then, from \cite{Rahn:02a} we obtain for the decoding matrix
\begin{equation}
\label{DET}
\mathscr{D}=\mathscr{E}^t \mathscr{T}.
\end{equation}
\begin{lemma}
\label{4vars}
The non-zero elements of $\mathscr{D}_X$  must be contained in $X_S$, and similarly for $Z$, although usually not for $Y$.
\end{lemma}
In particular this implies that if $\sigma=X$ or $\sigma=Z$, then $\mathscr{G}_{\sigma \sigma'}$ depends only on $N_{\sigma I}$, $N_{\sigma X}$, $N_{\sigma Y}$, and $N_{\sigma Z}$. Then to find convergence of the $X$ and $Z$ rows, we can look at these rows separately.

\begin{proof}
Since $\mathscr{D}_{I}=I$, the non identity stabilizer elements must commute with half of the recovery operators. Only the non-zero elements of $\mathscr{D}_\sigma$ don't commute with exactly half of the recovery operators. This implies that each non-identity element of $S(X)$ commutes with half of the elements of $R_Z=R(\epsilon_X,Z)$, and similarly for $S(Z)$ and $R_X=R(\epsilon_Z,X)$. If half of either $R_X$ or $R_Z$ commute with some element of $S$, then half of all of the the recovery operators  commute with it. Now, pick some non-zero element $c=X^{\otimes n} s_X s_Z$ of $\mathscr{E}_X$, where $s_i \in S(i)$. If $c \notin A_X$ then   $s_Z \neq I$. Then, if an element $r \in A_X$, it follows that $\eta(r,c)=\eta(r,s_Z)$, and so, half of $R_X$ commutes with $c$. Then, $c$ must commute with half of the recovery elements, and so must be zero in $\mathscr{D}_{X}$. Then the non-zero elements of $\mathscr{D}_X$ are in $A_X$.
\end{proof}

\begin{theorem}
\label{cssfuncs}
There exists functions $f_1(a,b,c,d)$ and $f_2(a,b,c,d)$ such that the following are is true for $\mathscr{G}=\Omega^c(\mathscr{N}^{\otimes n})$.
\begin{eqnarray*}
\mathscr{G}_{XI}=&f_1(\mathscr{N}_{XI},\mathscr{N}_{XX},i\mathscr{N}_{XY},\mathscr{N}_{XZ})\\
\mathscr{G}_{XX}=&f_2(\mathscr{N}_{XX},i\mathscr{N}_{XY},\mathscr{N}_{XZ},\mathscr{N}_{XI})\\
\mathscr{G}_{XY}=&i^n f_2(i\mathscr{N}_{XY},\mathscr{N}_{XZ},\mathscr{N}_{XX},\mathscr{N}_{XI})\\
\mathscr{G}_{XZ}=&f_2(\mathscr{N}_{XZ},\mathscr{N}_{XX},i\mathscr{N}_{XY},\mathscr{N}_{XI})\\
\mathscr{G}_{ZI}=&f_1(\mathscr{N}_{ZI},\mathscr{N}_{ZX},i\mathscr{N}_{ZY},\mathscr{N}_{ZZ})\\
\mathscr{G}_{ZX}=&f_2(\mathscr{N}_{ZX},i\mathscr{N}_{ZY},\mathscr{N}_{ZZ},\mathscr{N}_{ZI})\\
\mathscr{G}_{ZY}=&i^n f_2(i\mathscr{N}_{ZY},\mathscr{N}_{ZZ},\mathscr{N}_{ZX},\mathscr{N}_{ZI})\\
\mathscr{G}_{ZZ}=&f_2(\mathscr{N}_{ZZ},\mathscr{N}_{ZX},i\mathscr{N}_{ZY},\mathscr{N}_{ZI}).
\end{eqnarray*}
Furthermore these functions $f_1(a,b,c,d)$ and $f_2(a,b,c,d)$ are symmetric under permutations of $b$, $c$, and $d$.

\end{theorem}

\begin{proof}
The permutation $X \rightarrow  iY \rightarrow Z \rightarrow X$, sends $\mathscr{E}_I=S$ to itself, and sends
\begin{equation*}
\mathscr{E}_X \rightarrow i^n \mathscr{E}_Y \rightarrow \mathscr{E}_Z \rightarrow \mathscr{E}_X.
\end{equation*}
Then, from lemma \ref{4vars}, and the fact that $X \leftrightarrow Z$ sends $\mathscr{D}_X \leftrightarrow \mathscr{D}_Z$, $f_1$ and $f_2$ must exist as stated.

As for the symmetries, $\mathscr{G}_{XI}$ depends on $\mathscr{D}_X$ and $\mathscr{E}_I$. By permuting $X$, $iY$, and $Z$, we preserve the stabilizer elements which are the non-zero elements of $\mathscr{E}_I$, and so $\mathscr{G}_{XI}$ is fixed under permutations of $N_{XX}$, $iN_{XY}$, $N_{XZ}$. $\mathscr{G}_{XX}$ depends on $\mathscr{D}_X$, and $\mathscr{E}_X$. By permuting $I$, $Z$, and $iY$, we preserve the non-zero elements of $\mathscr{E}_X$, which are $\overline{X}$  times the elements of $\mathscr{E}_I$ (see  Eq. (\ref{encodedpauli})), and so $\mathscr{G}_{XX}$ is fixed under permutations of $N_{XI}$, $iN_{XY}$, and $N_{XZ}$. The other cases follow similarly.
\end{proof}

\begin{lemma}
\label{nosign}
Let $\sigma \neq \sigma'$ be single qubit Pauli matrices and let $\sigma''$ be a non-zero element of $\mathscr{E}_\sigma$. Then $\sigma'$ appears tensored an even number of times in $\sigma''$.
\end{lemma}
\begin{proof}
In the case where $\sigma=I$, $\mathscr{E}_\sigma$ corresponds to the stabilizer group. Since $S$ is generated by even weight elements in $X_S$ and even weight elements in $Z_S$, in order for it to be Abelian, it must have the above property. For general $\sigma$ we have $\mathscr{E}_\sigma=\overline{\sigma} S$, and, using $\overline{\sigma}$ is $\sigma$ on all qubits, the desired result follows.
\end{proof}

\begin{theorem}
\label{cssconverge}
A CSS code takes a channel $\mathscr{N}$ to the identity channel if and only if both  vectors $[\mathscr{N}_{XI},\mathscr{N}_{XX},\mathscr{N}_{XZ},i\mathscr{N}_{XY}]$ and $[\mathscr{N}_{ZI},\mathscr{N}_{ZZ},\mathscr{N}_{ZX},i\mathscr{N}_{ZY}]$
converge to $[0,1,0,0]$ under the map

\begin{eqnarray*}
[a,b,c,d] \to \cr
[f_1(a,b,c,d),f_2(b,c,d,a),f_2(c,d,a,b),if_2(d,a,b,c)].
\end{eqnarray*}

In fact, it is sufficient that they converge to $[*,1,*,*]$.

\end{theorem}
\begin{proof}
Obviously, this is a necessary condition.
From Lemma \ref{nosign}, we see that each of the variables $b$, $c$, and $d$ in $f_1(a,b,c,d)$, and $f_2(a,b,c,d)$ must appear an even number of times in each term. So we may ignore any $-1$ sign in front of $\mathscr{G}_{XY}$ or $\mathscr{G}_{ZY}$. From the symmetries we have, it then follows that the above map determines convergence on the $X$ and $Z$ rows. The rest of the theorem follows from Thm. \ref{altconv}.
\end{proof}

\noindent
{\bf Remark (Unital channels):}
In the case of unital channels, the above reduces to the condition that both
$[\mathscr{N}_{XX},\mathscr{N}_{XZ},i\mathscr{N}_{XY}]$ and $[\mathscr{N}_{ZZ},\mathscr{N}_{ZX},i\mathscr{N}_{ZY}]$
converge to $[1,*,*]$ under the map
\begin{equation*}
[a,b,c] \to [f_2(a,b,ic,0),f_2(b,ic,a,0),if_2(ic,a,b,0)].
\end{equation*}
Notice that this no longer depends on $f_1$.

\begin{lemma}
\label{nondiagtermcss}
For CSS codes, we have
$\max(c_d,c_m) \leq 2^{\frac{3}{2} (n-k)}$
for $c_d$ and $c_m$ as defined in Thm. \ref{nondiagterm}.
\end{lemma}
\begin{proof}
We use the bound of Thm. \ref{nondiagterm} for the non-diagonal terms. In the case of a CSS code, we have for $A=X$ or $A=Z$ that $\mathscr{D}_A \subset A_S$, and so the non-zero entries are given by $S(A) A^{\otimes n}$. Therefore the sum in Eq. (\ref{coffboundseq}) has only $2^\frac{n-k}{2}$ entries, giving  an overall coefficient of $2^{\frac{3}{2} (n-k)}$.
\end{proof}

\subsubsection{Doubly-even CSS codes}
Doubly even CSS codes are CSS codes that have weight divisible by $4$ for $S(X)$ and $S(Z)$. For these codes we can strengthen Thm. \ref{cssconverge}. Define functions $g_1$ and $g_2$ that are the same as the $f_1$ and $f_2$ defined in Thm. \ref{cssfuncs}, without the factors of $i$.
\begin{theorem}
\label{doublyconverge}
A doubly even CSS code takes a channel $\mathscr{N}$ to the identity channel if an only if both $[\mathscr{N}_{XI},\mathscr{N}_{XX},\mathscr{N}_{XZ},\mathscr{N}_{XY}]$ and $[\mathscr{N}_{ZI},\mathscr{N}_{ZZ},\mathscr{N}_{ZX},\mathscr{N}_{ZY}]$
converge to $[0,1,0,0]$ under the map
\begin{eqnarray*}
[a,b,c,d] \to \cr
[g_1(a,b,c,d),g_2(b,c,d,a),g_2(c,d,a,b),g_2(d,a,b,c)].
\end{eqnarray*}
\end{theorem}
\begin{proof}
The stabilizer group is formed by generators $\in X_S$, and generators $\in Z_S$, each with weight divisible by 4. Then $X$ and $Z$ together appear a number of times divisible by $4$  in each stabilizer element (and similarly for $\{X,Y\}$, $\{Y,Z\}$). Following similar reasoning to that of the proof of lemma \ref{nosign}, we find that $c$ and $t$ together appear a divisible by $4$ number of times in each term of $f_j$ ($j=1,2$). Then, $f_j(a,b,c,id)=f_j(a,b,ic,d)$, and by definition
\begin{equation}
\label{fg}
g_j(a,b,c,d)=f_j(a,b,c,id).
\end{equation}
These $g_j$ satisfy all the symmetries above and the convergence relations of Thm. \ref{cssconverge} (without the factors of $i$).
\end{proof}

\subsubsection{Example: $[[7,1,3]]$ CSS code}
We use the example of the $[[7,1,3]]$ code, a doubly even CSS code commonly used in fault tolerance calculations, to illustrate how to find the functions defined in Thm. \ref{cssfuncs} and use Thm.  \ref{cssconverge} to analyze the convergence of channels under this code.

\paragraph{Computation of the coding map} The stabilizer group of this code is generated by the elements $IIIXXXX, IXXIIXX, XIXIXIX$ and $IIIZZZZ, IZZIIZZ, ZIZIZIZ$.
Using the notation from section \ref{another}, the non-zero elements of  $\mathscr{E}_I$ are the stabilizer group elements.
\begin{align*}
\mathscr{E}_I =  \sum_{s \in S} s = & (IIIIIII+IIIXXXX)\\(IIIIIII+IXXIIXX)&(IIIIIII+XIXIXIX) \\
+ &(IIIIIII+IIIZZZZ)\\(IIIIIII+IZZIIZZ)&(IIIIIII+ZIZIZIZ)
\end{align*}
We have $\overline{X}=XXXXXXX$, and $\overline{Z}=ZZZZZZZ$.
One notices that there are $7$ terms that are some permutation of $IIIXXXX$. Let $p_7(IIIXXXX)$ denote the sum over these permutations. $p_7(IIIYYYY)$ and $p_7(IIIZZZZ)$ give us the corresponding permutations of $IIIYYYY$ and $IIIZZZZ$. Similarly, there are $42$ terms that are $-IZZXXYY$, up to some permutation, so we define a function $p_{42}(-IZZXXYY)$ to sum over these.
Then we can write
\begin{eqnarray*}
\mathscr{E}_I = IIIIIII +  p_{42} (-IZZXXYY)\\
+p_7(IIIXXXX+IIIYYYY+IIIZZZZ).
\end{eqnarray*}
With  $\mathscr{E}_\sigma = \mathscr{E}_I \overline{\sigma}$ we get
\begin{align*}
\mathscr{E}_X = XXXXXXX + p_{42} (-XYYIIZZ)\\ + p_7(XXXIIII+XXXZZZZ+XXXYYYY) \\
\mathscr{E}_Y = -YYYYYYY - p_{42} (-YXXZZII)\\ - p_7(YYYZZZZ+YYYIIII+YYYXXXX) \nonumber\\
\mathscr{E}_Z = ZZZZZZZ + p_{42} (-ZIIYYXX)\\ + p_7(ZZZYYYY+ZZZXXXX+ZZZIIII).\nonumber
\end{align*}
The recovery operators which depend on $X$ are
\begin{align*}
R(\varepsilon_Z,X)=\{IIIIIII,XIIIIII,IXIIIII,\\IIXIIII,IIIXIII,IIIIXII,IIIIIIXI,IIIIIIX\}.\nonumber
\end{align*}
Combining these with the recovery operations in $R(\varepsilon_X,Z)$, we easily find all $64$ recovery operators. There are $1$ in the form $IIIIIII$, all $7$ permutations of $IIIIIIX$, all $7$ permutations of $IIIIIIY$, all $7$ permutations of $IIIIIIZ$, and all $42$ permutations of $IIIIIXZ$. Eq. (\ref{DET}) now allows us to find the elements of $\mathscr{D}_{\sigma}$.
We calculate $\mathscr{D}_X$ from $\mathscr{E}_X$. $XXXXXXX$ commutes with $\frac{8}{64}$ recovery elements, $XXXIIII$ commutes with $\frac{40}{64}$ recovery elements, and $XXXZZZZ$, $XXXYYYY$, and $XYYIIZZ$ each commute with $\frac{32}{64}$ of the recovery elements. Then
\begin{eqnarray*}
\mathscr{D}_X=&\frac{1}{4}p_7(XXXIIII) &-\frac{3}{4}XXXXXXX\\
=&\frac{1}{4}XXXIIII&+\frac{1}{4}XIIXXII\\+&\frac{1}{4}IXIXIXI&+\frac{1}{4}IIXIXX\nonumber\\
+&\frac{1}{4}IIXXIIX&+\frac{1}{4}IXIIXIX\\+&\frac{1}{4}XIIIIXX&-\frac{3}{4}XXXXXXX\nonumber.
\end{eqnarray*}
A similar calculation shows that $\mathscr{D}_Z = -\frac{3}{4}ZZZZZZZ+\frac{1}{4}p_7(ZZZIIII)$, but $\mathscr{D}_{Y}$ doesn't follow this pattern.

Now we wish to compute $\mathscr{G}_{XI}$. First we look at how the  $\frac{1}{4}p_7(XXXIIII)$ component of $\mathscr{D}_{X}$ contributes. From $N_{I \sigma}=\delta_{I \sigma}$, it follows that only elements in $\mathscr{D}_I$ that are identity on the last $4$ qubits contribute. This is just $IIIIIII$, so we get $p_7(\frac{1}{4}N_{XI}N_{XI}N_{XI}N_{II}N_{II}N_{II}N_{II})=\frac{7}{4}N_{XI}^3$.
For the $-\frac{3}{4}XXXXXXX$ component of $\mathscr{D}_X$, everything in $\mathscr{E}_I$ contributes. This gives a contribution of
\begin{align*}
&-\frac{3}{4}(N_{XI}N_{XI}N_{XI}N_{XI}N_{XI}N_{XI}N_{XI}\\
&+p_7(N_{XI}N_{XI}N_{XI}N_{XX}N_{XX}N_{XX}N_{XX}\nonumber\\
& +N_{XI}N_{XI}N_{XI}N_{XY}N_{XY}N_{XY}N_{XY}\\
&+N_{XI}N_{XI}N_{XI}N_{XZ}N_{XZ}N_{XZ}N_{XZ})\\
& +p_{42}(-N_{XI}N_{XZ}N_{XZ}N_{XX}N_{XX}N_{XY}N_{XY})).\nonumber
\end{align*}
Together, these give
\begin{eqnarray*}
\mathscr{G}_{XI}=\frac{7}{4}N_{XI}^3-\frac{3}{4}(N_{XI}^7+7N_{XI}^3(N_{XX}^4+N_{XY}^4\\+N_{XZ}^4)-42N_{XI}N_{XX}^2N_{XY}^2N_{XZ}^2).
\end{eqnarray*}
A similar calculation shows that
\begin{eqnarray*}
\mathscr{G}_{XX}=\frac{7}{4}N_{XX}^3-\frac{3}{4}(N_{XX}^7+7N_{XX}^3(N_{XI}^4+N_{XZ}^4\\+N_{XY}^4)-42N_{XX}N_{XI}^2N_{XZ}^2N_{XY}^2)\\
-\mathscr{G}_{XY}=\frac{7}{4}N_{XY}^3-\frac{3}{4}(N_{XY}^7+7N_{XY}^3(N_{XI}^4+N_{XX}^4\\+N_{XZ}^4)-42N_{XY}N_{XI}^2N_{XX}^2N_{XZ}^2)\nonumber\\
\mathscr{G}_{XZ}=\frac{7}{4}N_{XZ}^3-\frac{3}{4}(N_{XZ}^7+7N_{XZ}^3(N_{XI}^4+N_{XY}^4\\+N_{XX}^4)-42N_{XZ}N_{XI}^2N_{XY}^2N_{XX}^2).\nonumber
\end{eqnarray*}
For the functions $g_j$, which are related to $f_j$ by Eq. (\ref{fg}), we obtain
\begin{eqnarray*}
g(a,b,c,d):=g_1(a,b,c,d)=g_2(a,b,c,d)\\=\frac{7}{4}a^3-\frac{3}{4}a^7-\frac{21}{4}a^3(b^4+c^4+d^4)+\frac{63}{2}ab^2c^2d^2.
\end{eqnarray*}
Note that
\begin{eqnarray*}
\mathscr{G}_{XI}&=&g(N_{XI},N_{XX},N_{XY},N_{XZ})\\
\mathscr{G}_{XX}&=&g(N_{XX},N_{XY},N_{XZ},N_{XI})\nonumber\\
\mathscr{G}_{XY}&=&-g(N_{XY},N_{XZ},N_{XI},N_{XX})\nonumber\\
\mathscr{G}_{XZ}&=&g(N_{XZ},N_{XI},N_{XX},N_{XY}).\nonumber
\end{eqnarray*}

\paragraph{Analysis}
We consider the convergence of a row of the channel matrix $[a,b,c,d]$ as in Thm. \ref{doublyconverge}.
We have from Thm. \ref{rownorm} that
\begin{equation}
\label{norm1}
a^2+b^2+c^2+d^2 \leq 1.
\end{equation}
If the channel is diagonal (or in general in the case where all but one parameter $a,b,c$ or $d$ are zero) we have a critical point $x_c=0.870807$ such that $g(\pm x_c,0,0,0)=\pm x_c$.

Let us now analyze the behavior of non-diagonal channels with small off-diagonal elements.
\begin{theorem}\label{lastthm}
If any of $a$, $b$, $c$, or $d$ is within $x_c$ of $0$, it must go to $0$.
\end{theorem}
\begin{proof}
This can be proved in general by a rather lengthy calculation. To convey the main idea we will here only give the proof in the case where one of the $4$ variables equals $0$ (for example, a unital channel). Then our function $g$ becomes $g(a,b,c)=\frac{7}{4}a^3-\frac{3}{4}a^3(a^4+7b^4+7c^4)$. We want to show that that if $0 < |a| < x_c$, we have that $|a|> |g(a,b,c)|$. Without loss of generality, we may assume that $a$ is positive. Below the critical value $x_c$, we have $a > \frac{7}{4}a^3-\frac{3}{4}a^7$, so we only need to see if $a \leq -g(a,b,c)$, which is maximized by $b=\sqrt{1-a^2}$, $c=0$. A simple calculation shows that there is no solution. Then it follows that $|a|$ must monotonically go to $0$.
\end{proof}
 From Thm. \ref{lastthm} and Eq.  (\ref{norm1}) we easily see that the vector $[a,b,c,d]$ must converge to a vector with at most one non-zero coefficient. Now suppose that $a$ is slightly above $x_c$, and that $b$, $c$, and $d$ have absolute values of at most some small $\epsilon$. We wish to see how much $\epsilon$ changes the critical convergence value for $a$. Let $k:=\frac{dg(a,0,0,0)}{da}|_{x_c}=1.691859$. Then,
\begin{eqnarray*}
g(a,b,c,d) \geq g(a,\epsilon,\epsilon,\epsilon) \\ \geq g(a,0,0,0)-\frac{63}{4}a^3\epsilon^4 \approx  k(a-x_c)+x_c-\frac{63}{4}a^3\epsilon^4.
\end{eqnarray*}
Since $b$, $c$, and $d$ become $O(\epsilon^d)=O(\epsilon^3)$ up to 4th order of $\epsilon$, the vector converges  to $[1,0,0,0]$ for $k(a-x_c)+x_c-\frac{63}{4}a^3\epsilon^4 \geq x_c$, which implies that
\begin{equation*}
\label{thisequation}
k(a-x_c) \geq \frac{63}{4}a^3\epsilon^4 \approx \frac{63}{4}\epsilon^4(x_c^3+3(a-x_c)^2) \approx \frac{63}{4}\epsilon^4x_c^3.
\end{equation*}
Solving up to first order for our new critical value, we get
\begin{equation*}
a=\frac{63x_c^3\epsilon^4}{4k}+x_c=6.14726\epsilon^4+x_c.
\end{equation*}
This implies that the off-diagonal terms affect the threshold to fourth order (as implied by Thm. \ref{nondiagterm}); but here we improved the prefactor $c_t$. Note that Lemma \ref{nondiagtermcss} would have given a prefactor of $512$.

If we choose a larger number instead of $6.14726$, for example $7$, then our vector converges to $[1,0,0,0]$ from $[x_c+7\epsilon^4,\epsilon,\epsilon,\epsilon]$ for $\epsilon$ as big as $0.3$.

\ignore{
If we would like that the vector converges {\em monotonically} to $[1,0,0,0]$, then
Eq. (\ref{thisequation}) becomes
\begin{equation*}
k(a-x_c) \geq \frac{63}{4}a^3\epsilon^4 +(a-x_c)\approx \frac{63}{4}\epsilon^4x_c^3 + (a-x_c),
\end{equation*}
which gives
\begin{equation*}
a=\frac{63x_c^3\epsilon^4}{4(k-1)}+x_c=15.03241\epsilon^4+x_c=c\epsilon^4+x_c.
\end{equation*}
The actual results here are significantly better than those we would obtain from Thm. \ref{nondiagterm} or Lemma \ref{nondiagtermcss}.
}

\section{SVD canonical form}
\label{sec:canon}
In this section, we follow the method of \cite{king:01}, applying unitary gates before and after our channel to create a new channel that has fewer parameters. This can be used to improve the region of convergence to the identity channel.

\begin{lemma}
\label{so3}
Let $\sigma_j$ be the non-identity elements of the Pauli group $\mathscr{P}$. Then if $U=e^{i \frac{\theta}{2} \sigma_1}$, then the unitary channel $\rho \rightarrow U \rho U^\dagger$ performs a rotation by $\theta$ in the $\sigma_3 \sigma_2$ plane. Expressing the unitary gates as channels in the Pauli basis creates a bijection from $SU(2) / (\pm I)$ to $1 \oplus SO(3)$.
\end{lemma}

The Singular Value decomposition (SVD) theorem \cite{Bhatia} states that if $A$ is a real matrix, then there exists $D=O_2^\dagger A O_1$ such that the $O_i$ are orthogonal, and $D$ is a diagonal matrix with elements $\lambda_i \geq 0$, which are called the singular values of $A$. Then $D = \sgn(\det A) R_2^\dagger A R_1^\dagger$, where $R_i \in SO(n)$.

\begin{theorem}
If $\mathscr{N}^{(1)}$ is a channel on one qubit, then there exists a channel
\begin{equation}
\label{canon}
\mathscr{T} = \mathscr{U}_2^{\dagger} \mathscr{N}^{(1)} \mathscr{U}_1 = \bordermatrix{\cr
\cr
&1 & 0 & 0 & 0 \cr
&t'_1 & \pm \lambda_1 & 0 & 0 \cr
&t'_2 & 0 & \pm \lambda_2 & 0 \cr
&t'_3 & 0 & 0 & \pm \lambda_3 \cr
},
\end{equation}
where $\mathscr{U}_i \in SU(2)$, and the $\pm$ designates the sign of $\det \mathscr{N}^{(1)}$.
\end{theorem}
\begin{proof}
 From Eq. \ref{noise}, define the vector $\mathbf{t} = (N_{XI},N_{YI},N_{ZI})$, and let $A$ be the $3 \times 3$ matrix with the other $9$ variable elements. From the SVD theorem, we have
\begin{equation*}
\mathscr{T} =\bordermatrix{\cr
\cr
& 1 & 0 \cr
& 0 & R_2^\dagger \cr
}\bordermatrix{\cr
\cr
& 1 & 0 \cr
& \mathbf{t} & A \cr
}\bordermatrix{\cr
\cr
& 1 & 0 \cr
& 0 & R_1\cr
} =\bordermatrix{\cr
\cr
& 1 & 0 \cr
& \mathbf{t'} & \pm D \cr
},
\end{equation*}
where $\mathbf{t'}=R_2^\dagger \mathbf{t}=(t'_1,t'_2,t'_3)$. The outer matrices are unitary channels by lemma \ref{so3}.
\end{proof}
Note that $\norm{\mathbf{t}}=\norm{\mathbf{t'}}$, so if the channel is unital, $\mathbf{t'}=\mathbf{0}$.
\subsection{CSS codes}
We now apply the above to CSS codes, and in particular examine the $[[7,1,3]]$ CSS code.
\begin{proposition}
For a given CSS code with a channel $\mathscr{T}$ in the canonical form of Eq. \ref{canon}, if at least 2 of $[t'_1, \lambda_i]$ converge to $[0,1]$ under the map $[a,b] \to [f_1(a,b,0,0),f_2(b,a,0,0)]$, where the $f_i$ are the functions from Thm. \ref{cssfuncs}, then by applying unitary gates before and after the channel $\mathscr{T}$, we can create a new channel  $\mathscr{T}'$ that converges to the identity.
\end{proposition}
\begin{proof}
Suppose that $[t'_i,\lambda_i]$ and $[t'_j, \lambda_j]$ converge to $[0,1]$ under the given map. We define a matrix $A \in SO(4)$ such that $\sigma_I \to \sigma_I$, $\sigma_i \to \sigma_X$, $\sigma_j \to \sigma_Z$, and the diagonal matrix $B=[1, \pm 1, 1, \pm 1]$. By lemma \ref{so3}, these are unitary channels.  Then,

\begin{equation*}
 \mathscr{T}'=A\mathscr{T}A^\dagger B = \bordermatrix{\cr
\cr
&1 & 0 & 0 & 0 \cr
&t'_i & \lambda_i & 0 & 0 \cr
&* & 0 & * & 0 \cr
&t'_j & 0 & 0 & \lambda_j \cr
},
\end{equation*}
and the rest follows from Thm. \ref{cssconverge}

\end{proof}

Note that even if $\mathscr{T}'$ was diagonal, the order of the $\lambda_i$ could affect whether it converges to the identity channel.

\subsubsection{Example: The $[[7,1,3]]$ CSS code}

For the $[[7,1,3]]$ code, the map is $h([a,b]) = [g(a,b),g(b,a)]$, where $g(a,b)=\frac{a^3}{4} (7-3a^4-21b^4)$. Let $[a_n,b_n] = h^{\circ k}([a,b])$. This converges to $[0,1]$ if and only if $b_n \to 1$.

Now we are interested in which $[a,b]$ converge to $[0,1]$. Using that $a^2 + b^2 \leq 1$, a numerical calculation shows that there is always convergence to $[0,1]$ for  $b > b_c \approx 0.927334$. For $[a,b]=[\sin \theta, \cos \theta]$, this threshold is exact, and so these converge to $[0,1]$ for $\abs{\theta} < \theta_c \approx 0.383572$. For a unital channel, $a=0$, and so this converges to $[0,1]$ for $b > x_c \approx 0.870807$. In either of these cases, we just need at most one singular value of the channel to be less than or equal to the given critical value.

We can find an approximate solution for the region of convergence to $[0,1]$ by solving $b_n \geq x_c$. For $n=1$, we have an approximation for the region of  $a^4 \leq f_1(b)=\frac{1}{3} - \frac{1}{7}b^4 - \frac{4 x_c}{21 b^3}$. As $n$ increases, these approximate regions rapidly converge to the actual region of convergence to $[0,1]$. \enote{ Take out? Suppose we want to find the magnitude of error in $a^4$ from using this calculation with a fixed $n$. For large $n$, $a_n \approx \frac{7}{4}a_{n-1}^3 (1-3b_n^4) \approx \frac{7}{4}a_{n-1}^3 (1-3 x_c^4)  \approx (-1)^n (1.126 a)^{3^n}$. Suppose $b_n = x_c + \epsilon$. Then, $b_{n+1} \approx x_c + k\epsilon -\frac{21}{4}b_n^3 a_n^4$, then $\epsilon \approx  \frac{21}{4k}x_c^3 a_n^4 \approx 2a_n^4 = O(1.126 a)^{3^n 4}$. For $n=3$, this is less than $10^{-11}$ for $b > x_c$.}

\ignore{
The error in $a^4$ is roughly $l(b_n-x_c)$, where $l = \frac{df(b)}{db}|_{x_c} \approx 0.49$.
$a_n \approx a^{3^n}$. For $b < b_c$,
For large $n$, $a^n \approx \frac{7}{4} a_{n-1}^3 (1-3 x_c^4) \approx -c^2 a_{n-1}^3 \approx \frac{(1)^n}{c}(ca)^{3^n}$, where $c \approx 1.13$.
The error is roughly $a_n^4 = O((ca)^{3^n}) \approx (cl(b-x_c))^{3^n}$.

 We look at the convergence of this 2 dimensional space.
We converge for $[0,b]$ with $b > x_c$. To first order
$b_1\geq x_c$, and so $a^4\leq \frac{1}{3} - \frac{1}{7}b^4 - \frac{4 x_c}{21b^3}$.

A calculation using the iterative map shows that since $a^2 + b^2 \leq 1$, we always converge for $a > a_c = 0.927334$. Channels of the form $[\sin u, \cos u]$ have exactly this threshold.
The treshold is exactly that for  in other words, $[\sin u, \cos u]$, converges for $\abs{u} < 0.383571$. For $a < a_c$,
}
The singular values of a unitary channel are always $1$. Note that if the unitary channel from lemma \ref{so3} is in its original non canonical form, it converge to the identity channel for $\abs{\theta} < \theta_c$.
\section{Conclusion and further questions}
\label{sec:conclude}

\subsection{Drawbacks of our approach}
The approach of integrating the sequence of concatenated encoding and noise as a rather simple map from channels to channels is very powerful. By abstracting away from the details of the encoding and the noise process, it drastically reduces the number of parameters, and makes the coding process amenable to a dynamical systems type analysis. However, this approach sometimes comes at a price. By ignoring the details of the coding and correction process, we might get error thresholds above the actual thresholds if we accounted for all these details. The following example illustrates this, introducing the notion of a recovery function.

Suppose we have a $[[n,k,d]]$ stabilizer code. We define a recovery or error correcting function $R(\varepsilon)$ \cite{JT} which maps the collection of syndromes measured by the codes to some $n$ qubit Pauli operator, $R : \mathscr{F}_{2^{n-k}} \to \mathscr{P}_n$. We also define a syndrome function $\varepsilon : \mathscr{P}_n \to \mathscr{F}_{2^{n-k}}$, which maps Pauli errors to some syndrome. With these definitions we must have that $\beta=\varepsilon(R(\beta))$, for any $\beta \in \mathscr{P}_n$. Note that we can chose $R(\beta)$ up to elements of the stabiliser $S$ without any difference for error correction. Hence our choices for $R(\beta)$ differ from each other by elements of the centralizer $C(S)$ are limited to the $4^k$ elements of the Centralizer modulo the Stabilizer. They can be written as an element of $C(S)$ times some representative element of $S$. To study the choice of recovery function on the channel, define
the matrix $T^{\sigma} \in \mathscr{M}_{4^n, 4^n}$ to be the diagonal matrix
\begin{equation*}
T^{\sigma}_{\sigma' \sigma'}=\frac{1}{2^{n-k}}\eta(\sigma, \sigma').
\end{equation*}
Then the matrix operator ${\cal T}$, defined in Eq. (\ref{eq:recover}), is ${\cal T}=\sum_i T^{R_i}$. We have $\mathscr{G}=\sum_i \mathscr{G}^{R_i}=\Omega^C(\mathscr{N})$, where the quasi-channel (they don't have to preserve trace)
\begin{equation*}
\mathscr{G}^{R_i}=\mathscr{E}^t \mathscr{T}^{R_i}\mathscr{N}\mathscr{E},
\end{equation*}
is the contribution of a single $R_i$ on the channel map.

When we measure a syndrome $\varepsilon$ during error-correction, we gain some information about the channel. Let the encoded state be described by the density matrix $\rho = \rho_I I+ \rho_X X + \rho_Y Y+ \rho_Z Z$. We can re-write our channel $\mathscr{G}=\Omega^C(\mathscr{N})$ as a sum over all syndromes
\begin{equation*}
\mathscr{G}'=\sum_{\beta \in F_{2^{n-k}}} \mathscr{G}^{R(\beta)} \otimes \ket{\beta}.
\end{equation*}
If we measure $\ket{\beta}$ and use the information, we collapse to a syndrome $\beta$ with probability $p_\beta=tr(\mathscr{G}^{R(\beta)} \rho) = 2\sum_{\sigma} \mathscr{G}^{R(\beta)}_{I\sigma } \rho_\sigma$, and the resulting density matrix is $\frac{1}{p_{\beta}} \mathscr{G}^{R(\beta)} \rho$. In particular, if $ \mathscr{G}_{IX} ^{R(\beta)}=\mathscr{G}_{IY} ^{R(\beta)}=\mathscr{G}_{IZ} ^{R(\beta)}=0$, then $p_\beta=
2\mathscr{G}^{R(\beta)}_{II}\rho_I=\mathscr{G}^{R(\beta)}_{II}$, which doesn't depend on $\rho$, and the resulting $\rho$-independent channel is then $\frac{1}{p_\beta}\mathscr{G}^{R(\beta)}$. If we throw this information away we recover the coding map $\mathscr{G}$ from the previous sections. In other words the coding map approach corresponds to ignoring the information about the channel that  we could have obtained  from the syndrome measurements, to optimize the recovery functions.

By performing measurements on the subblocks of a concatenated code, we affect the channel on each qubit of the top level code. If we don't optimize our error correction, we are not being as efficient as we should be. For example, a distance $3$ code can't correct some $2$ qubit errors, and so the code we obtain by concatenating it once with itself without changing the error correction function can't fix some $4$ qubit errors. However, the distance $d$ of a distance $d_1$ code concatented with a distance $d_2$ code is $d \geq d_1 d_2$, and so we should be able to correct any $4$ qubit error. The problem is to keep track of all of this syndrome information, and finding the optimal error correction function seems to be computationally hard.

\ignore{
\paragraph{Example:}
Let $[1,x,x,x]$ be a diagonal channel, and let us look  at the channel map of the $[[5,1,3]]$ code \cite{Rahn:02a}. It has $\Omega^{Five}([1,x,x,x])=[1,g_1(x),g_1(x),g_1(x)]$, where $g_1(x)= \frac{5}{2}x^3-\frac{3}{2}x^5$. The part ${\cal G}^{I}$ resulting from the identity syndrome is $\frac{1}{16}[1+15x^4,g_2(x),g_2(x),g_2(x)]$, where $g_2(x)=10x^3+6x^5$. Then, if on every block of every level of concatenation of this code, we keep measuring the $0$ syndrome (this event happens of course with probability tending to zero), we get a resulting channel map $[1,f(x),f(x),f(x)]$ with $f(x)=\frac{10x^3+6x^5}{1+15x^4}$. Then the channel converges to identity if $x > \frac{1}{3}$, which is much better than in general for this code (compare with  \jnote{insert number} from \cite{Rahn:02a}). We note that $f(x)$ is not a polynomial \jnote{Jesse more}. A similiar calculation shows that for each of the other syndromes, we get $\frac{1}{16}[1-x^4,g_3(x),g_3(x),g_3(x)]$, where $g_3(x)=2x^3-2x^5$. Then if we never get the $0$ syndrome (this also happens with probability tending to zero), we get $f(x)=\frac{2x^3}{1+x^2}$, which only corrects if $x=1$.
}

\subsection{Open questions}
We have initiated a dynamical systems approach to quantum error correction, extending the result of Rahn et al. \cite{Rahn:02a}. This only opens the road to further analysis and many questions remain open. We list a few of them here.

In our analysis we have always assumed that an error correction process is successful, if the associated coding map takes the noise channel to the identity channel. However, this might be too stringent a condition.
Are there any other criteria for information retrieval, which are not equivalent to zero (corrected) error?

Another question relates to the basin of correctable noise for a code: If our noise channel lies outside the basin of attraction of a
  certain code, can we find another code that would ``lift'' this
  noise into the basin of attraction of the old code? More
  specifically, given a code $C$ (with $d \geq 3$) and a noise channel
  $p \in \Delta - \mathscr{B}_C$, is there another code $C'$ such that
  $\Omega^{C'}(p) \in \mathscr{B}_C$? If the answer is positive, then
  the concatenation scheme $C^k \circ C'$ corrects $p$, as $k
  \rightarrow \infty$. It would be interesting to formalise these ideas.

Yet another question concerns the shape of the region of correctable noise. Is there a (non-trivial) bound for the size or shape of the
  domain of attraction? Can we characterize regions of noise that are
  not correctable by any code? There is a new and interesting bound on
  noise from which no circuit can recover in \cite{Razborov:03a}.
  However the methods used there are not dynamical. Is it possible to make
  sharper statements?

\section*{Acknowledgment}

The authors would like to thank Birgitta Whaley  for support and fruitful conversations. JF and JK acknowledge support by DARPA and Air Force
Laboratory, Air Force Material Command, USAF, under agreement number
F30602-01-2-0524, and by DARPA and the Office of Naval Research under
grant number FDN-00014-01-1-0826. JF, JK and SNS are partially supported by NSF ITR grant CCF-0205641. JK is supported by ACI S\'ecurit\'e Informatique,
2003-n24, projet ``R\'eseaux Quantiques", ACI-CR 2002-40 and EU 5th
framework program RESQ IST-2001-37559.


%

\end{document}